\documentclass[amsmath,amssymb,aps, prl,twocolumn, superscriptaddress]{revtex4-2}
\usepackage{graphicx} 
\usepackage[colorlinks=true,linkcolor=blue,citecolor=blue, urlcolor=blue]{hyperref}
\usepackage[all]{hypcap} 

\begin{document}

\title{Quantum Linear Magnetoresistance: A Modern Perspective}
\author{Shuai Li}
\email{lis1277@ctgu.edu.cn}
\affiliation{Hubei Engineering Research Center of Weak Magnetic-field Detection, Department of Physics, China Three Gorges University, Yichang 443002, China}
\author{Huichao Wang}
\email{wanghch26@mail.sysu.edu.cn}
\affiliation{Guangdong Provincial Key Laboratory of Magnetoelectric Physics and Devices, School of Physics, Sun Yet-sen University, Guangzhou, China.}
\begin{abstract}
Magnetoresistance is a powerful probe for characterizing the intrinsic physics embedded in materials. Among its various manifestations, linear magnetoresistance has a long history and continues attracting research interest. In contemporary studies, a clear understanding of the magnetoresistance character of quantum origin is more crucial than ever for the study of  emerging materials. In this perspective, we examine the linear magnetoresistance of quantum mechanism, from its theoretical basis to experimental studies, and discuss open questions and promising future research directions in this field.
\end{abstract}

\maketitle

Magnetoresistance (MR) is the resistance change of materials under an applied magnetic field. When the applied magnetic field becomes strong, the electronic energy band of the material quantizes into discrete Landau levels in the plane perpendicular to the field direction. As the magnetic field strength increases, the degeneracy of Landau levels also increases, making the number of occupied Landau level drop. This process manifests as oscillations in MR with the magnetic field strength, a phenomenon known as the Shubnikov–de Haas (SdH) effect. The oscillations stop when only the lowest Landau level remains occupied, that is, the system enters the quantum limit. MR that originates from the Landau level quantization of the system is known as the quantum MR.

In the early study of MR, analysis from the classical kinetic equation showed that, in systems with closed Fermi surface, MR increases quadratically with magnetic field until it reaches saturation. This result aligns with many experimental observations. However, starting in 1928, Kapitza reported that, in some materials \cite{Kapitza_PRSA_1928,Kapitza_PRSA_1929}, the MR was linearly dependent on magnetic field strength, and the saturation was not observed. This phenomenon became known as linear MR (LMR). Later, LMR in polycrystalline samples of metals was well explained by the averaging effect in systems with open Fermi surface \cite{Lifshitz_Sov.Phys.–JETP_1959}. In 1997, LMR was reported in doped silver chalcogenides \cite{Xu_N_1997}, sparking a renewed research interest. Two famous theoretical explanations for LMR emerged in this period. One is the quantum-mechanism LMR proposed by Abrikosov at the microscopic level \cite{Abrikosov_PRB_1998}, in which the linear electronic dispersion and screened-Coulomb type impurity were considered. The other is the resistor-network model proposed by Parish and Littlewood at the macroscopic level \cite{Parish_N_2003}. More recently, beginning in 2013 \cite{Wang_PRB_2013}, the surge in research on topological materials revitalized interest in LMR. This is because the LMR derived by Abrikosov is based on linear dispersion, a characteristic shared by most topological materials. Lately, the observation of LMR in graphene \cite{Xin_N_2023} and tellurium \cite{Tang2025} has further promoted the study of LMR related to electron-electron and electron-phonon interactions.

Based on the theoretical approaches employed, explanations for LMR can be generally classified into three types: classical, semiclassical, and quantum mechanisms, as shown in Fig.~\ref{Fig}(a-c). A representative classical theory is the aforementioned resistor-network model. LMR in this model is actually contributed by the Hall components due to the carrier concentration fluctuation \cite{Parish_N_2003,Kisslinger_PRB_2017}. Another frequently cited classical mechanism is the LMR in finite-size samples near charge neutrality \cite{Alekseev_PRL_2015}. These classical LMR mechanisms are based on classical Drude transport theory. Consequently, the resulting LMR is not tied to the microscopic properties of the material; instead, the size and geometry of the sample can influence the result. For the semiclassical mechanism, a well-known example is the LMR caused by guiding center diffusion in the smooth potential \cite{Song_PRB_2015}. The semiclassical theories have not fully considered the Landau quantization; thus they are widely applied to LMR in materials under weak magnetic fields. Theories of LMR based on the quantum transport formalism are distinctly different from those of classical or semiclassical. Instead of being applicable to various systems, the quantum LMR is closely related to the band structure and impurity potential. In addition, currently, all the LMR from a full quantum method are derived in the quantum limit.

\section{Theoretical Considerations}

\begin{figure*}[t]
\centering
\includegraphics{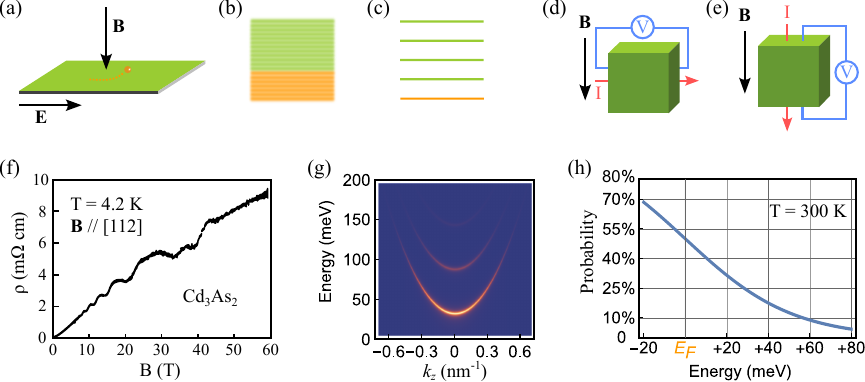}
\caption{Sketches for the LMR of (a) classical, (b) semiclassical, and (c) quantum mechanisms. In (b) and (c), the green color denotes unoccupied states and the orange color denotes occupied states. In the classical theory, the Lorentz force governs the motion of charged particles. In semiclassical case, energy dispersion is considered, but the Landau levels are squeezing together. In the quantum theory, with a fully quantum mechanical description, the Landau levels are clearly resolved, and the quantum limit is considered. (d) and (e) are sketches of the experimental setups used to measure the transverse MR and longitudinal MR, respectively. (f) LMR observed in$\mathrm{Cd_3As_2}$ \cite{Zhao_PRX_2015}. The MR not only exhibits a linear behavior in the quantum limit regime but also has a linear background in the SdH oscillation regime. (g) Spectral density of the 3D electron gas under a z-directional magnetic field. The lightness (color gradient from red to orange to white) denotes the carrier occupation. Parameters used in the plot are: the effective mass of 0.1 times the free electron mass, magnetic field strength of 50 T, Fermi energy of 20 meV, and temperature of 300 K. (h) Plot of Fermi-Dirac distribution function at 300 K.}
\label{Fig}
\end{figure*}

LMR of the quantum mechanism is often derived within the framework of linear response theory.  While the Kubo formula for conductivities used by different authors is often written in different forms, they all can be derived from 
\begin{equation}
\sigma_{\alpha\alpha}=\frac{\pi\hbar e^{2}}{V}\sum_{u,u'}\int\frac{\partial n\left(\omega\right)}{\partial\omega}A_{u}\left(\omega\right)A_{u'}\left(\omega\right)\left|\left\langle u'\right|v_{\alpha}\left|u\right\rangle \right|^{2}d\omega, \label{Eq:sigma}    
\end{equation}
where $\hbar$, $e$, and $V$ are the reduced Planck constant, elementary charge, and volume of the system, respectively. The Fermi-Dirac distribution function is denoted by $n\left(\omega\right)$. The Dirac notation $\left|u\right\rangle$ represents the state with quantum number $u$, and the corresponding spectral function is denoted by $A_{u}$. The velocity operator along $\alpha$ direction is $v_{\alpha}$. Equation~(\ref{Eq:sigma}) is the general formula for both transverse and longitudinal conductivities. For systems in a strong magnetic field, the quantum number $u$ includes the wavevector parallel to the field direction, indices for Landau degeneracy, Landau level, spin and orbital. The integral over the energy $\omega$ together with the partial derivative of $n(\omega)$ indicate that, in low temperatures, the conductivity is primarily contributed by carriers near the Fermi energy. From this expression, one can find that the field $\mathbf{B}$ always appears in the form of absolute value. For instance, the summation over the Landau degeneracy index results in a pre-factor proportional to the strength of the magnetic field, regardless of its direction. Therefore, the conductivity found from this formula always satisfies the property, $\sigma_{\alpha\alpha}(\mathbf{B})=\sigma_{\alpha\alpha}(\mathbf{-B})$, i.e. the Onsager's relation holds for quantum LMR.

The field dependence of MR derived from the Kubo formula is critically dependent on the microscopic electronic properties, as explicitly shown in Eq.~(\ref{Eq:sigma}). The spectral function $A$, eigenvector $\left|u\right\rangle$, and velocity operator $v_a$ are all derived directly from the electronic Hamiltonian. Consequently, a deviation from the specific band structure required by a quantum-mechanism LMR theory can cause the MR to deviate from linearity. The aforementioned quantum LMR derived in Abrikosov's work \cite{Abrikosov_PRB_1998} is for three-dimensional (3D) systems with a linear dispersion, and it does not hold under the same deduction routine when the system reduces to two-dimensional (2D). Compared to 3D systems, in 2D systems, there is no wavevector along the magnetic field direction, and the integral in Eq.~(\ref{Eq:sigma}) reduces to the 2D case. These mathematical differences lead to distinct results. Furthermore, in materials, additional bands near the Fermi energy that are not considered in the quantum LMR theory can also affect the result. In conclusion, quantum LMR has strict conditions, but this makes it useful for probing the microscopic electronic properties of materials.

The field dependence of carrier scattering time is also a key factor that determines the field dependence of MR. Different types of scattering potential result in distinct field-dependent scattering times, which in turn affect the field dependence of MR. Therefore, the quantum LMR is not exclusively tied to linear dispersion; it can also exist in systems with parabolic dispersion provided they have specific scattering potentials. Known quantum LMR can arise from weak impurity potentials of screened-Coulomb or long-range Gaussian type \cite{Li_PRB_2023}, strong delta-type impurity potentials \cite{Abrikosov_JPAMG_2003}, and acoustic-phonon scattering \cite{Arora_PRB_1977}. Therefore, to understand the underlying physics of quantum LMR observed in a specific material, it's crucial to determine which scattering potential dominates. In Eq. (\ref{Eq:sigma}), all scattering effects are included in the spectral function $A(\omega)$ via self-energy. Typically, the first Born approximation (often shortened as the Born approximation) is used for self-energy, which is valid for impurities of weak strength and low concentration. However, the LMR derived from the quantum method also has a dependence on the approximation method used. For example, the T-matrix approximation (sometimes referred to as the full Born approximation) for self-energy is required for impurities of high strength; under this consideration, it has been shown that quantum LMR can exist in systems with parabolic dispersion and delta-type potential impurities \cite{Abrikosov_JPAMG_2003}. Furthermore, Eq. (\ref{Eq:sigma}) corresponds to the one-loop Feynman diagram, and corrections may be needed in the calculation. The well-known vertex correction for conductivity calculations, in the case of impurity scattering, weights the scattering events and yields the transport time. Its importance has been demonstrated in the longitudinal MR of the single-node massless Weyl mode \cite{Li_PRB_2023}. 

In the literature, the LMR usually refers to the phenomenon in transverse MR, which is measured in setups where the excitation current is perpendicular to the magnetic field direction and the voltage is measured along the direction of excitation current [Fig.~\ref{Fig}(d)]. In 2D systems, the magnetic field is typically applied in the out-of-plane direction, and MR is measured in the plane. In 3D systems, the longitudinal MR is usually also measured, in setups where the excitation current, voltage measurement and magnetic field are all parallel [Fig.~\ref{Fig}(e)]. Since the quantum method provides a unified theoretical framework for both transverse and longitudinal MR, the presence of LMR in one direction must be accompanied by a specific field-dependent MR in the other direction. For example, in Abrikosov's model, it has been shown that the transverse LMR is accompanied by a longitudinal MR that is inversely dependent on the field strength \cite{Li_PRB_2023}. Therefore, when comparing the experimental measurements with theoretical results, it is best to analyze both transverse and longitudinal MR to clearly identify the underlying quantum mechanism. It is worth noting that, in theoretical studies, the transverse LMR is usually found in models that are isotropic in the plane perpendicular to field direction, and the Hall conductivity in these models is inversely in proportion to the field strength. Consequently, with $\rho_{xx(yy)}=\sigma_{xx}/(\sigma_{xx}^2+\sigma_{xy}^2)$, the LMR is found when the transverse conductivity is also inversely proportional to the field strength. On the one hand, in real materials, it is possible for highly anisotropic systems to have LMR in one direction and another type of field-dependent MR in different directions within the same plane perpendicular to the field. As MR is derived from the conductivity in theoretical study (the resistivity tensor is related to the conductivity tensor through matrix inverse operation), the expression of transverse resistivity in the anisotropic system becomes the complex:  $\rho_{xx(yy)}=\sigma_{yy(xx)}/(\sigma_{xx}\sigma_{yy}-\sigma_{xy}\sigma_{yx})$. This complexity makes calculating the LMR in anisotropic systems more challenging than in isotropic systems. A careful theoretical investigation into the LMR in highly anisotropic systems is currently lacking. On the other hand, it has been found that in some materials, the Hall resistance is not linear in field strength. Whether a LMR can arise in this kind of system remains to be explored.

The quantum LMR generally accompanies the quantum limit of the system. Yet, no single LMR theory holds across the semiclassical, quantum oscillation, and quantum limit regimes. However, it has been reported that in some materials, the MR oscillates on a linearly field-dependent background, and develops into LMR once the quantum limit is reached. A typical example is seen in $\mathrm{Cd_3As_2}$, as shown in Fig.~\ref{Fig}(f) \cite{Zhao_PRX_2015}. Since LMR is observed in the quantum limit, classical and semiclassical theories are not suitable for explaining this experimental observation. Nevertheless, a concrete quantum theory that accounts for both the linear background of the SdH oscillation and the linear MR in quantum limit is currently lacking. It has been shown that the MR background in the quantum oscillation regime in Abrikosov's model is proportional to $B^{1/3}$, and it is only possible to achieve a linear background in SdH oscillation by using an artificial scattering rate proportional to $B$  \cite{Xiao_PRB_2017, Koenye_PRB_2018}. In addition, a partially quantum approach has shown that LMR can exist across the semi-classical regime to the quantum oscillation regime in the topological surface state with Zeeman splitting \cite{Wang_PRB_2012}.

Generally, a high magnetic field is required to reach the quantum limit. High field strength can produce large gaps between Landau levels, ensuring they are well separated, and a large degeneracy of Landau levels, allowing the lowest Landau level to have enough states for all carriers. In high quality samples with a low carrier concentration, however, the quantum limit can be achieved even under low magnetic fields. Under this circumstance, the gaps between Landau levels are small. Consequently, high temperature can lead to a finite occupation of carriers on the Landau levels higher than the lowest one. As demonstrated in Fig.~\ref{Fig}(g), for a 3D electron gas model under magnetic field along z-direction, a high temperature leads to a tail-like carrier occupation that spreads to Landau bands higher than the lowest one. When the magnetic field strength increases, the Landau degeneracy increase, and the gaps between Landau levels become greater, resulting in this occupation tail smoothly shrinking until all carriers are on the lowest Landau level. Since there is no abrupt change of Landau level occupation during this process, no MR oscillations are exhibited. This phenomenon is therefore distinct from the case of multiple Landau level occupation at low temperatures, which supports the SdH oscillation with increasing field strength. This temperature-induced occupation tail exists in recent LMR experiments in high-mobility graphene \cite{Xin_N_2023} and tellurium \cite{Tang2025}. In these experiments, the LMR starts at a field strength of a few tesla at room temperature. Hence, a claim of high-temperature quantum LMR from the lowest Landau level (or band) requires a rigorous evaluation showing the dominance of the occupation on the lowest Landau level (or band). It is worth noting that even when the energy gaps between Landau levels are much greater than the room-temperature $k_BT$ (25.85 meV), a rigorous evaluation through the Fermi-Dirac distribution function indicates that a small number of carriers still occupy the higher Landau levels, as shown in Fig.~\ref{Fig}(h). For the case of a long ``carrier occupation tail'' on higher Landau levels, its effect on the quantum LMR awaits a detailed theoretical study.

\section{Experimental Considerations}

In experimental studies, claiming the observation of quantum LMR must be done with great caution. Systematic measurements and thorough analysis are necessary to confirm its quantum origin and distinguish it from diverse classical or semiclassical mechanisms. Any discussion of quantum LMR should begin with a rigorous assessment of sample quality. Careful attention must be given to sample geometry, including shape and thickness. Variations in thickness can cause differences in Hall voltage at different points in the sample, which in turn appear as additional resistance along its length, producing LMR \cite{Bruls_PRL_1981, Pippard1989}. This effect is more pronounced when thickness changes are significant, or when the sample is relatively wide. If the width approaches the length, uneven current flow can become substantial, leading to LMR induced by transverse Hall voltages \cite{Parish_N_2003}. Techniques such as X-ray diffraction, scanning electron microscopy, and transmission electron microscopy can be used to examine the sample’s microstructural features. Carrier concentration and mobility are key physical parameters. Hall measurements should be performed, at different positions on the sample whenever possible, to detect potential variations in carrier concentration or mobility—common causes of LMR through classical mechanisms that must be ruled out. High-quality materials are essential for achieving quantum LMR, which depends on well-resolved Landau levels made possible by high mobility. Observing consistent LMR behavior across multiple high-quality crystalline samples supports a quantum origin by ruling out explanations based on specific defects or random factors.

Systematic measurements and careful tuning can provide substantial evidence for identifying quantum LMR. In the measurement setup, the symmetry and placement of electrodes can influence the results and should be designed with care. Adjusting electrode positions and comparing results from two-probe and four-probe methods can help determine whether the observed LMR comes from intrinsic behavior or from artificial effects caused by current distribution \cite{Delmo_N_2009}. The temperature dependence of LMR is critical and should be studied over a wide range to understand the conditions under which it appears. For instance, the quantum LMR in tellurium caused by phonon scattering tends to occur at higher temperatures rather than at low temperatures, with the LMR slope showing an inverse $1/T$ dependence on temperature \cite{Tang2025}. In contrast, most quantum LMR models are based on impurity scattering, requiring low temperatures to ensure impurity scattering dominates over phonon scattering. In this case, the slope is mainly determined by impurity concentration \cite{Abrikosov_PRB_1998}. Some classical mechanisms, such as LMR from inhomogeneity, can persist across both low and high temperatures \cite{Kisslinger_NP_2015, Narayanan_PRL_2015}, even maintaining a constant slope over a wide temperature range \cite{Khouri_PRL_2016}. Another important factor is to carefully examine the magnetic field range in which LMR appears. Quantum LMR emerges when the sample reaches the quantum limit, which typically requires a relatively high magnetic field. This contrasts with classical mechanisms, where LMR can appear even at very low magnetic fields within the classical regime \cite{Khouri_PRL_2016}. Accurate estimation of the quantum limit field is key to determining whether LMR is tied to quantum origin. Additionally, multi-parameter transport measurements beyond MR are important. For instance, analyzing Hall resistivity in strong magnetic fields can offer deep insights into carrier properties and the Fermi surface, which are valuable for distinguishing different origins.

Finally, it is important to emphasize that claiming the observation of quantum LMR requires rigorous data analysis, not just a visual check of whether the $R(B)$ curve appears straight. It is recommended to present $dR/dB$ vs. B or $d\rho/dB$ vs. $B$ plots, or to use quantitative approaches such as power-law fitting to verify LMR \cite{Tang2025}. A linear curve shows a constant differential, which is more sensitive and objective than simply judging the linearity of the $R–B$ curve by eye. This method also allows for a clear definition of the magnetic field range in which LMR begins and ends.


\emph{ACKNOWLEDGEMENTS}
This work is financially supported by the National Natural Science Foundation of China (Grant Nos. 12504059, 21BAA01133, 12374052, 92165204), Hubei Provincial Natural Science Foundation of China (Grant No. 2024AFB289), Guangdong Basic and Applied Basic Research Foundation (Grant No. 2023A1515010487), Guangdong Provincial Quantum Science Strategic Initiative (Grant No. GDZX2401009), Guangzhou Basic and Applied Basic Research Foundation (Grant No. 2025A04J5405) and Guangdong Provincial Key Laboratory of Magnetoelectric Physics and Devices (Grant No. 2022B1212010008).

\bibliography{LMR-perspective}

\end{document}